\documentclass[conference]{IEEEtran}
\IEEEoverridecommandlockouts

\usepackage{cite}
\usepackage{amsmath,amssymb,amsfonts}
\usepackage{algorithmic}
\usepackage{graphicx}
\usepackage{textcomp}
\usepackage{xcolor}
\usepackage{booktabs}
\usepackage{subcaption}
\usepackage[linesnumbered,vlined,ruled]{algorithm2e}
\usepackage[hidelinks]{hyperref}
\usepackage{balance}

\newtheorem{remark}{Remark}

\newcommand{\rev}[1]{{\color{blue}#1}}

\renewcommand{\rev}[1]{#1}

\newcommand{\revcolor}{}

\def\BibTeX{{\rm B\kern-.05em{\sc i\kern-.025em b}\kern-.08em
    T\kern-.1667em\lower.7ex\hbox{E}\kern-.125emX}}
\begin{document}

\title{Deep Reinforcement Learning for BD-RIS-Assisted Multi-UAV RSMA Communications}

\author{\IEEEauthorblockN{Aamer Mohamed Huroon\IEEEauthorrefmark{1}, Ali A. Nasir\IEEEauthorrefmark{1}\IEEEauthorrefmark{2}, and Daniel B. da Costa\IEEEauthorrefmark{1}\IEEEauthorrefmark{2}}
\IEEEauthorrefmark{1}Interdisciplinary Research Center for Communication Systems and Sensing (IRC-CSS)\\
King Fahd University of Petroleum and Minerals (KFUPM), Dhahran, Saudi Arabia\\
\IEEEauthorrefmark{2} Department of Electrical
Engineering, King Fahd University of Petroleum and Minerals,\\ Dhahran, Saudi Arabia.\\
Email: \{aamer.mohamed, anasir, daniel.costa\}@kfupm.edu.sa}

\maketitle

\begin{abstract}
Beyond-diagonal reconfigurable intelligent surfaces (BD-RISs) extend conventional diagonal RISs by allowing controllable coupling among reflecting elements. This flexibility is attractive for aerial networks, where unmanned aerial vehicles (UAVs) face rapidly varying channels and strict power limits. In this paper, we study a hybrid BD-RIS-assisted multi-UAV rate-splitting multiple access (RSMA) downlink, in which UAVs within favorable BD-RIS coverage use both direct and BD-RIS-aided links, whereas the remaining UAVs are served through direct links only. We formulate a network sum-rate maximization problem that jointly optimizes UAV trajectories, RSMA common-rate allocation, beamforming, BD-RIS scattering matrices, and UAV-BD-RIS association. \rev{The resulting mixed-integer non-convex problem is solved by using deep deterministic policy gradient (DDPG) algorithm, equipped with a constraint-preserving action layer that generates reciprocal and lossless BD-RIS responses in closed form and handles the binary association inside the environment transition. Simulation results, averaged over three training seeds under a physically consistent channel model with blocked direct links near the BD-RISs, show that the proposed design attains a sum rate comparable to SAC and higher than twin delayed deep deterministic policy gradient ( TD3), proximal policy optimization (PPO), and deep Q-network (DQN), exceeds per-slot alternating optimization with fixed trajectories at about $1/65$ of its per-slot runtime, and benefits mainly from RSMA, while the additional BD-RIS gain is modest.}
\end{abstract}


\section{Introduction}

Reconfigurable intelligent surfaces (RISs) shape the wireless propagation environment via almost passive reflecting elements. In conventional RIS designs, the reflection matrix is diagonal and each element is controlled individually. Beyond-diagonal RIS (BD-RIS) architectures relax this restriction by introducing reconfigurable inter-element coupling, which can redistribute incident signals across surface ports and improve array gain, interference suppression, and spatial multiplexing \cite{li2022bd,li2026tutorial,nerini2026graph,wu2026physics}. \rev{BD-RISs have also been investigated for integrated sensing and communication (ISAC) and non-terrestrial networks (NTNs) \cite{magbool2025survey,zhang2025sagin,isacbdzheng2025,khan2026ntn}.} These features are particularly relevant to UAV networks, where mobility, altitude-dependent line-of-sight probability, and limited onboard power make link adaptation challenging \cite{granelli2026uav,mao2026uav}.

Rate-splitting multiple access (RSMA) is another flexible technology for interference-limited multiuser transmission. RSMA divides user messages into common and private parts, partially decodes interference and treats the rest as noise, thereby bridging space-division multiple access (SDMA) and non-orthogonal multiple access \cite{mao2022rsma,aboumahmoud2024rsma,khoshafa2025rsma}. RSMA handles intra-cluster interference at the information-theoretic layer, while BD-RIS manipulates propagation at the electromagnetic layer.

The joint optimization of BD-RIS-aided UAV RSMA communications networks is challenging. The beamformers, common-rate allocation, BD-RIS scattering matrices, and UAV-surface associations are strongly coupled; BD-RIS constraints are matrix-valued; and UAV channels are time-varying. Existing works have studied aerial RIS communications with conventional optimization \cite{huroon2024gbd,cantos2026mimo} and learning-based approaches \cite{huroon2025madr,huang2026iot}, but a hybrid direct/BD-RIS multi-UAV RSMA model with a learning-driven solution remains relatively unexplored. We emphasise that the primary contribution of this work is the system level design, the joint model and the problem-specific learning design, and using reinforcement-learning (RL) based algorithm. The main contributions can be summarised as follows:
\begin{enumerate}
\item We propose a hybrid BD-RIS-assisted multi-UAV RSMA downlink architecture, where UAVs in favorable propagation regions use both direct and BD-RIS-assisted links, while other UAVs rely on direct air-to-ground links, and RSMA manages intra-cluster interference.

\item We formulate a network sum-rate maximization problem that jointly optimizes RSMA common-rate allocation, active beamforming, BD-RIS scattering matrices, UAV trajectories, and UAV-BD-RIS association, resulting in a strongly coupled mixed-integer non-convex problem.

\item \rev{We solve the problem by using the deep deterministic policy gradient (DDPG) algorithm, tailored through a constraint-preserving action layer, a closed-form reciprocal and lossless BD-RIS response, and an environment-side treatment of the binary association. Results averaged over multiple seeds and compared with the twin delayed deep deterministic policy gradient ( TD3), soft actor–critic (SAC), proximal policy optimization (PPO), deep Q-network (DQN) and alternating optimization (AO).}
\end{enumerate}

\section{System Model}

\begin{figure}[t]
    \centering
    \includegraphics[width=0.86\linewidth]{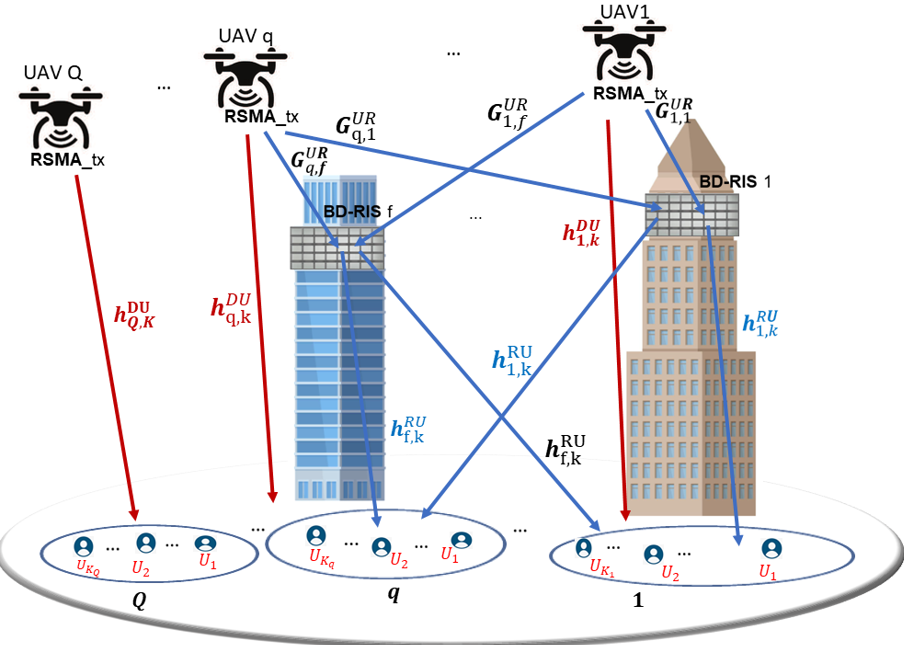}
    \caption{BD-RIS-assisted multi-UAV RSMA communication system.}
    \label{fig:system_model}
\end{figure}

As illustrated in Fig.~\ref{fig:system_model}, we consider a downlink network with $Q$ UAV transmitters, $F$ BD-RISs, and multiple ground-user clusters. The UAV and BD-RIS sets are $\mathcal{Q}=\{1,\ldots,Q\}$ and $\mathcal{F}=\{1,\ldots,F\}$. UAV $q\in\mathcal{Q}$ serves a dedicated cluster $\mathcal{K}_q=\{1,\ldots,K_q\}$ of single-antenna users. Each UAV has $N$ transmit antennas and each BD-RIS has $M$ passive scattering elements.

The BD-RIS deployment is \emph{hybrid}: some UAVs lie in favorable propagation regions of the BD-RISs and exploit the cascaded paths, while others are far from all BD-RISs, so their cascaded links suffer severe path loss and are not used. The association variable $b_{q,f}\in\{0,1\}$ indicates whether UAV $q$ is associated with BD-RIS $f$, and $a_q=\min(1,\sum_{f\in\mathcal{F}} b_{q,f})$ indicates whether UAV $q$ is assisted by at least one BD-RIS. The assisted and direct-only sets are $\mathcal{Q}_R=\{q: a_q=1\}$ and $\mathcal{Q}_D=\{q: a_q=0\}$.

Orthogonal time/frequency resource blocks are allocated to the UAVs, with fraction $\tau_q\in(0,1]$ for UAV $q$ and $\sum_{q}\tau_q\le 1$; equal allocation $\tau_q=1/Q$ is used unless otherwise stated. Hence, there is no inter-UAV interference, and the only interference is the intra-cluster interference among the RSMA streams.

The mission horizon is divided into $T$ slots, $\mathcal{T}=\{1,\ldots,T\}$. Let $\mathbf{u}_q[t]=[x_q[t],y_q[t],H_q[t]]^T$ be the position of UAV $q$ at slot $t$, and $\mathbf{U}=\{\mathbf{u}_q[t]\}_{q,t}$ the network trajectory. The mobility constraints are
\begin{align}
\footnotesize
&\|\mathbf{u}_q[t+1]-\mathbf{u}_q[t]\|\le V_q^{\max}\Delta_t,\quad \forall q,t<T, \label{eq:speed}\\
&H_q^{\min}\le H_q[t]\le H_q^{\max},\quad \forall q,t, \label{eq:altitude}
\end{align}
where $V_q^{\max}$ is the maximum speed and $\Delta_t$ is the slot duration. Since the channels depend on $\mathbf{u}_q[t]$, the trajectory is jointly optimized with beamforming, BD-RIS scattering, and RSMA rate allocation.

\subsection{RSMA Transmit Signal}

For UAV $q$, the message of user $k\in\mathcal{K}_q$ is split into a common and a private part. The common parts are jointly encoded into $s_{q,c}$, and the private parts into $\{s_{q,k}\}_{k\in\mathcal{K}_q}$, with unit-power streams. The transmitted signal is
\begin{equation}
\mathbf{x}_q[t]=\mathbf{w}_{q,c}[t]s_{q,c}[t]+\sum_{k\in\mathcal{K}_q}\mathbf{w}_{q,k}[t]s_{q,k}[t],
\label{eq:tx_signal}
\end{equation}
where $\mathbf{w}_{q,c}[t],\mathbf{w}_{q,k}[t]\in\mathbb{C}^{N\times1}$ are the common and private beamformers, subject to
\begin{equation}
\footnotesize
\|\mathbf{w}_{q,c}[t]\|^2+\sum_{k\in\mathcal{K}_q}\|\mathbf{w}_{q,k}[t]\|^2\le P_q^{\max},\quad \forall q,t.
\label{eq:power}
\end{equation}

\subsection{Channel Model and BD-RIS Architecture}

The direct UAV-user channel is $\mathbf{h}_{q,k}^{\mathrm{DU}}\in\mathbb{C}^{1\times N}$, the UAV-BD-RIS channel is $\mathbf{G}_{q,f}^{\mathrm{UR}}\in\mathbb{C}^{M\times N}$, and the BD-RIS-user channel is $\mathbf{h}_{f,k}^{\mathrm{RU}}\in\mathbb{C}^{1\times M}$. The UAV-BD-RIS and BD-RIS-user links, with the BD-RISs mounted on building facades, follow Rician fading, i.e., for a generic channel $\mathbf{H}$,
\begin{equation}
\mathbf{H}=\sqrt{\beta(d)}\Big(\sqrt{\tfrac{\kappa}{\kappa+1}}\mathbf{H}_{\mathrm{LoS}}+\sqrt{\tfrac{1}{\kappa+1}}\mathbf{H}_{\mathrm{NLoS}}\Big),
\end{equation}
where $\beta(d)=\beta_0 d^{-\alpha}$ is the path loss at distance $d$, $\beta_0$ is the path loss at 1~m, $\alpha$ is the path-loss exponent, and $\kappa$ is the Rician factor. \rev{$\mathbf{H}_{\mathrm{LoS}}$ is built from the array-response vectors of the half-wavelength uniform linear array (ULA) at each UAV and the uniform planar array (UPA) at each BD-RIS, and the entries of $\mathbf{H}_{\mathrm{NLoS}}$ are independent $\mathcal{CN}(0,1)$. The direct UAV-user links follow the elevation-dependent LoS model \cite{alhourani2014optimal}: the link is LoS (Rician, as above) with probability $p_{\mathrm{LoS}}(\vartheta)=1/(1+a\,e^{-b(\vartheta-a)})$, where $\vartheta$ is the elevation angle in degrees, and otherwise NLoS with an excess loss $\eta_{\mathrm{NLoS}}$. Clusters located next to a BD-RIS are in a dense high-rise area, where the direct links are frequently blocked, whereas the other clusters are in an open area (parameters in Table~\ref{tab:params}).}

\rev{We adopt a \emph{fully connected} BD-RIS \cite{shen2022modeling,li2022bd}. Unlike a diagonal RIS, its off-diagonal entries let the signal incident on one element be redistributed to other elements before reradiation. Its scattering matrix $\boldsymbol{\Phi}_f\in\mathbb{C}^{M\times M}$ must satisfy}
{\revcolor
\begin{equation}
\footnotesize
\boldsymbol{\Phi}_f=\boldsymbol{\Phi}_f^{T},\qquad \boldsymbol{\Phi}_f^{H}\boldsymbol{\Phi}_f=\mathbf{I}_M,\quad \forall f\in\mathcal{F}.
\label{eq:passive}
\end{equation}}
\rev{Following the multiport network model \cite{shen2022modeling}, $\boldsymbol{\Phi}_f$ is generated from a real symmetric susceptance matrix $\boldsymbol{\Psi}_f=\boldsymbol{\Psi}_f^{T}\in\mathbb{R}^{M\times M}$ of the reconfigurable impedance network as}
{\revcolor
\begin{equation}
\footnotesize
\boldsymbol{\Phi}_f=(\mathbf{I}_M+jZ_0\boldsymbol{\Psi}_f)^{-1}(\mathbf{I}_M-jZ_0\boldsymbol{\Psi}_f),
\label{eq:cayley}
\end{equation}}
\rev{where $Z_0=50~\Omega$. Since $\mathbf{A}_f=jZ_0\boldsymbol{\Psi}_f$ is skew-Hermitian, \eqref{eq:cayley} is a Cayley transform and $\boldsymbol{\Phi}_f$ is unitary; since $\mathbf{A}_f$ is also symmetric, $(\mathbf{I}+\mathbf{A}_f)^{-1}$ and $(\mathbf{I}-\mathbf{A}_f)$ commute and $\boldsymbol{\Phi}_f^T=\boldsymbol{\Phi}_f$. Hence, every $\boldsymbol{\Psi}_f$ yields a hardware-realizable response satisfying \eqref{eq:passive}, so that the relaxed constraint $\boldsymbol{\Phi}_f^H\boldsymbol{\Phi}_f\preceq\mathbf{I}_M$, which may yield non-realizable responses, is not needed.}

Incorporating the association variables, the equivalent channel of user $k$ in cluster $q$ is
\begin{equation}
\footnotesize
\mathbf{h}_{q,k}[t]=\mathbf{h}_{q,k}^{\mathrm{DU}}[t]+
\sum_{f\in\mathcal{F}} b_{q,f}\,\mathbf{h}_{f,k}^{\mathrm{RU}}[t]\boldsymbol{\Phi}_f[t]\mathbf{G}_{q,f}^{\mathrm{UR}}[t],
\label{eq:eq_ch_with_assoc}
\end{equation}
which reduces to the direct channel $\mathbf{h}_{q,k}^{\mathrm{DU}}[t]$ for $q\in\mathcal{Q}_D$ (i.e., $b_{q,f}=0,\forall f$).

\subsection{Received Signal, SINRs, and Achievable Rates}

Since UAVs use orthogonal resources, user $k$ in cluster $q$ receives
\begin{equation}
y_{q,k}=\mathbf{h}_{q,k}\mathbf{w}_{q,c}s_{q,c}
+\mathbf{h}_{q,k}\mathbf{w}_{q,k}s_{q,k}
+\!\!\sum_{i\in\mathcal{K}_q, i\ne k}\!\!\mathbf{h}_{q,k}\mathbf{w}_{q,i}s_{q,i}
+n_{q,k},
\label{eq:rx_signal}
\end{equation}
where $n_{q,k}\sim\mathcal{CN}(0,\sigma_k^2)$; the three terms are the common stream, the desired private stream, and the intra-cluster interference. With one-layer RSMA, each user first decodes and removes the common stream and then decodes its private stream. For both assisted ($q\in\mathcal{Q}_R$) and direct-only ($q\in\mathcal{Q}_D$) clusters, the common- and private-stream SINRs are given by
\begin{align}
\footnotesize
\gamma_{q,k}^{c}[t]&=
\frac{|\mathbf{h}_{q,k}[t]\mathbf{w}_{q,c}[t]|^2}
{\sum_{i\in\mathcal{K}_q}|\mathbf{h}_{q,k}[t]\mathbf{w}_{q,i}[t]|^2+\sigma_k^2},
\label{eq:sinr_c}\\
\gamma_{q,k}^{p}[t]&=
\frac{|\mathbf{h}_{q,k}[t]\mathbf{w}_{q,k}[t]|^2}
{\sum_{i\in\mathcal{K}_q, i\ne k}|\mathbf{h}_{q,k}[t]\mathbf{w}_{q,i}[t]|^2+\sigma_k^2},
\label{eq:sinr_p}
\end{align}
where $\mathbf{h}_{q,k}[t]$ in \eqref{eq:eq_ch_with_assoc} automatically distinguishes the two cases. The common rate and the rate of user $k$ are
\begin{align}
\footnotesize
R_q^{c}[t]&=\min_{k\in\mathcal{K}_q}\log_2\big(1+\gamma_{q,k}^{c}[t]\big),\label{eq:rate_c}\\
R_{q,k}[t]&=C_{q,k}[t]+\log_2\big(1+\gamma_{q,k}^{p}[t]\big),\label{eq:rate_k}
\end{align}
where $C_{q,k}[t]\ge0$ is the common-rate portion of user $k$, satisfying
\begin{equation}
\footnotesize
\sum_{k\in\mathcal{K}_q}C_{q,k}[t]\le R_q^{c}[t],\quad \forall q,t.
\label{eq:common_alloc}
\end{equation}
The instantaneous and mission-averaged network sum rates are
\begin{equation}
\footnotesize
R_{\mathrm{sum}}[t]=\sum_{q\in\mathcal{Q}}\sum_{k\in\mathcal{K}_q}R_{q,k}[t],\qquad
\overline{R}_{\mathrm{sum}}=\frac{1}{T}\sum_{t\in\mathcal{T}}R_{\mathrm{sum}}[t].
\label{eq:sum_avg}
\end{equation}

\section{Problem Formulation}

We maximize the long-term network throughput by jointly optimizing the trajectories $\mathbf{U}$, common-rate allocation $\mathbf{C}$, beamforming $\mathbf{W}$, BD-RIS matrices $\boldsymbol{\Phi}$, and association $\mathbf{B}=\{b_{q,f}\}$:
\begin{subequations}
\footnotesize
\begin{align}
\mathbf{P1}:\ \max_{\mathbf{U},\mathbf{C},\mathbf{W},\boldsymbol{\Phi},\mathbf{B}}\ & \overline{R}_{\mathrm{sum}} \\
\mathrm{s.t.}\ \ & \eqref{eq:speed},\ \eqref{eq:altitude},\ \eqref{eq:power},\ \eqref{eq:passive},\ \eqref{eq:common_alloc}, \\
& R_{q,k}[t]\ge R_{q,k}^{\min},\quad \forall q,k,t, \\
& \textstyle\sum_{f\in\mathcal{F}} b_{q,f}\le F_q^{\max},\quad \forall q, \label{eq:Fmax}\\
& \textstyle\sum_{q\in\mathcal{Q}} b_{q,f}\le Q_f^{\max},\quad \forall f, \label{eq:Qmax}\\
& b_{q,f}\in\{0,1\},\quad \forall q,f,
\end{align}
\end{subequations}
where $R_{q,k}^{\min}$ is the QoS requirement, $F_q^{\max}$ is the maximum number of BD-RISs assisting UAV $q$, and $Q_f^{\max}$ is the maximum number of UAVs served by BD-RIS $f$. $\mathbf{P1}$ is non-convex because of the binary association, the trajectory-dependent channels, the coupled active/passive beamforming, the fractional SINRs, and the \rev{non-convex reciprocal and unitary} BD-RIS constraints.

\section{\texorpdfstring{\rev{Proposed DDPG-Based Solution}}{Proposed DDPG-Based Solution}}

\begin{figure}[t]
    \centering
    \includegraphics[width=0.8\linewidth]{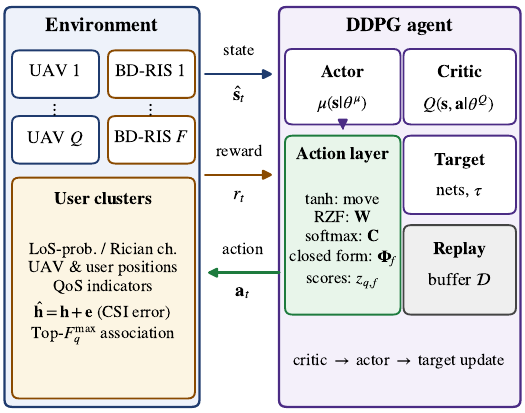}
    \caption{\rev{Proposed DDPG-based joint resource optimization framework.}}
    \label{fig:proposed_solution}
\end{figure}

Fig.~\ref{fig:proposed_solution} summarizes the learning framework. A central controller acts as the agent that observes the environment and selects the joint active/passive control actions. A deterministic policy suits the continuous, high-dimensional beamforming and scattering variables.

\begin{remark}
\rev{The learning engine is the standard DDPG algorithm, i.e., actor-critic networks, experience replay, soft target updates, Gaussian exploration noise, and deterministic policy gradients; we do not claim a new RL algorithm. The problem-specific contributions are the state/action design, the constraint-preserving action layer, the physically consistent BD-RIS parameterization \eqref{eq:cayley}, and the environment-side handling of the binary association.}
\end{remark}

\subsection{MDP Formulation}

At slot $t$, the state is
\begin{equation}
\footnotesize
\hat{\mathbf{s}}_t=\{\|\hat{\mathbf{h}}_{q,k}^{\mathrm{DU}}\|^2,\|\hat{\mathbf{G}}_{q,f}^{\mathrm{UR}}\|_F^2,\|\hat{\mathbf{h}}_{f,k}^{\mathrm{RU}}\|^2,\mathbf{u}_q,\mathbf{v}_{q,k},\mathbf{B}_{t-1},\boldsymbol{\rho}_t,P_q^{\max},t\},
\label{eq:state}
\end{equation}
where $\mathbf{v}_{q,k}$ is the user position, $\boldsymbol{\rho}_t$ collects QoS indicators, and $\hat{(\cdot)}$ denotes estimated CSI. \rev{The channel gains enter the state in dB, whereas the full estimated channels are used by the action layer $\Gamma(\cdot)$ below, which keeps the state dimension independent of $M$ and $N$.} \rev{To model imperfect CSI, every estimated channel is $\hat{\mathbf{h}}=\mathbf{h}+\mathbf{e}$, where $\mathbf{e}\sim\mathcal{CN}(\mathbf{0},\delta^2\beta(d)\mathbf{I})$ and $\delta^2$ is the normalized estimation-error variance. The agent decides based on $\hat{\mathbf{s}}_t$, whereas the achieved rates are computed with the true channels; $\delta^2=0$ recovers perfect CSI.}

\rev{\textit{Constraint-preserving action layer:} The actor output $\mathbf{a}_t\in[-1,1]^{D}$ (tanh output) is mapped by a fixed deterministic function $\Gamma(\cdot)$, which also uses the estimated CSI, to the physical controls $\{\boldsymbol{\Delta}_t,\mathbf{W}_t,\mathbf{C}_t,\boldsymbol{\Phi}_t,\mathbf{z}_t\}$, so that every continuous constraint holds by construction: i) the bounded outputs are scaled to UAV displacements with $\|\Delta\mathbf{u}_q[t]\|\le V_q^{\max}\Delta_t$, which satisfies \eqref{eq:speed}; ii) with $\hat{\mathbf{H}}_q$ stacking the estimated equivalent channels of cluster $q$, the private beams are the normalized columns of $\hat{\mathbf{H}}_q^H(\hat{\mathbf{H}}_q\hat{\mathbf{H}}_q^H+\nu_q\mathbf{I})^{-1}$, the common beam is the normalized $\hat{\mathbf{H}}_q^H\mathbf{c}_q$, where $\nu_q$ and $\mathbf{c}_q$ are actor outputs, and a softmax of $K_q+1$ outputs splits $P_q^{\max}$ among the streams, which satisfies \eqref{eq:power}; iii) the common rate is split as $C_{q,k}=\varpi_{q,k}R_q^c$, where $\{\varpi_{q,k}\}_k$ is a softmax output, which satisfies \eqref{eq:common_alloc}; and iv) for BD-RIS $f$ assisting UAV $q$, let $\mathbf{x}_f=\hat{\mathbf{G}}_{q,f}^{\mathrm{UR}}\mathbf{v}_q$ be the incident wavefront, with $\mathbf{v}_q$ the dominant right singular vector of $\hat{\mathbf{G}}_{q,f}^{\mathrm{UR}}$, and $\mathbf{y}_f=\sum_k \tilde{\varpi}_{q,k}e^{j\theta_k}(\hat{\mathbf{h}}_{f,k}^{\mathrm{RU}})^H/\|\hat{\mathbf{h}}_{f,k}^{\mathrm{RU}}\|$ the target wavefront, where $\{\tilde{\varpi}_{q,k}\}_k$ is a softmax output and $\theta_k=\angle(\hat{\mathbf{h}}_{q,k}^{\mathrm{DU}}\mathbf{v}_q)$ co-phases the reflected and direct signals. $\Gamma$ returns, in closed form, the symmetric unitary matrix with $\boldsymbol{\Phi}_f\bar{\mathbf{x}}_f=\bar{\mathbf{y}}_f$ ($\bar{\cdot}$: normalization), namely $\boldsymbol{\Phi}_f=\mathbf{V}_f^{*}\mathbf{S}_f\mathbf{V}_f^{H}+(\mathbf{V}_f^{\perp})^{*}(\mathbf{V}_f^{\perp})^{H}$, where $\mathbf{V}_f$ is an orthonormal basis of $\mathrm{span}\{\bar{\mathbf{x}}_f,\bar{\mathbf{y}}_f^{*}\}$, $\mathbf{V}_f^{\perp}$ completes it to a unitary matrix, and $\mathbf{S}_f$ is the $2\times2$ symmetric unitary matrix with $\mathbf{S}_f\mathbf{V}_f^H\bar{\mathbf{x}}_f=\mathbf{V}_f^T\bar{\mathbf{y}}_f$. This $\boldsymbol{\Phi}_f$ satisfies \eqref{eq:passive} exactly and is realized by \eqref{eq:cayley} with $\boldsymbol{\Psi}_f=(jZ_0)^{-1}(\mathbf{I}+\boldsymbol{\Phi}_f)^{-1}(\mathbf{I}-\boldsymbol{\Phi}_f)$, which is real and symmetric. Hence, the actor needs $K_q$ outputs per UAV instead of $M(M+1)/2$ per BD-RIS. No projection follows the actor, so the action produced, evaluated by the critic, and stored is identical, which avoids action-mismatch bias.}

\rev{\textit{Binary association:} The actor also outputs continuous scores $z_{q,f}$. The environment deterministically selects the $F_q^{\max}$ highest-scoring BD-RISs of each UAV and then keeps, for each BD-RIS, at most $Q_f^{\max}$ UAVs with the highest scores, so that \eqref{eq:Fmax}-\eqref{eq:Qmax} hold. This top-$k$ selection is part of the transition $P(\hat{\mathbf{s}}_{t+1}|\hat{\mathbf{s}}_t,\mathbf{a}_t)$, so the critic learns $Q(\hat{\mathbf{s}},\mathbf{a})$ with respect to the continuous scores and no gradient through the selection is required. Similarly, the altitude bound \eqref{eq:altitude}, the area boundary, and the reachability of the final UAV position are enforced by the environment through clipping.}

The reward balances spectral efficiency, QoS satisfaction, and constraint handling:
\begin{equation}
\footnotesize
r_t=R_{\mathrm{sum}}(t)-\eta_1\sum_{q,k}[R_{q,k}^{\min}-R_{q,k}(t)]^+-\eta_2\Omega_t,
\label{eq:reward}
\end{equation}
where $[x]^+=\max\{x,0\}$, \rev{$\Omega_t$ counts the corrections made by the environment (position clipping and association scores removed by the capacity limit $Q_f^{\max}$)}, and $\eta_1,\eta_2>0$ are penalty weights.

\begin{algorithm}[t]
\caption{\rev{DDPG for BD-RIS-Assisted Multi-UAV RSMA}}
\label{alg:ddpg}
\small
Initialize actor $\mu(\cdot|\theta^\mu)$, critic $Q(\cdot|\theta^Q)$, targets $\theta^{\mu'}\!\leftarrow\!\theta^\mu$, $\theta^{Q'}\!\leftarrow\!\theta^Q$, and replay buffer $\mathcal{D}$\;
\For{episode $=1,\ldots,E$}{
Initialize UAV/user positions and channels\;
\For{slot $t=1,\ldots,T$}{
Observe $\hat{\mathbf{s}}_t$ and select \rev{$\mathbf{a}_t=\mathrm{clip}\big(\mu(\hat{\mathbf{s}}_t|\theta^\mu)+\mathbf{n}_t,-1,1\big)$}, $\mathbf{n}_t\sim\mathcal{N}(\mathbf{0},\sigma^2\mathbf{I})$\;
\rev{Map $\mathbf{a}_t$ through $\Gamma(\cdot)$; the environment selects $\mathbf{B}_t$ from $\mathbf{z}_t$}\;
Execute, compute rates with true CSI, $r_t$, and $\hat{\mathbf{s}}_{t+1}$\;
Store $(\hat{\mathbf{s}}_t,\mathbf{a}_t,r_t,\hat{\mathbf{s}}_{t+1})$ in $\mathcal{D}$ \rev{(the same $\mathbf{a}_t$)}\;
Sample a minibatch of size $B$; update the critic by \eqref{eq:critic} and the actor by \eqref{eq:actor}\;
Soft-update: $\theta^{Q'}\!\leftarrow\!\tau\theta^Q\!+\!(1\!-\!\tau)\theta^{Q'}$, $\theta^{\mu'}\!\leftarrow\!\tau\theta^\mu\!+\!(1\!-\!\tau)\theta^{\mu'}$\;
}
}
\end{algorithm}

\subsection{Training Procedure}

The actor $\mu(\mathbf{s}|\theta^\mu)$ outputs deterministic actions, and the critic $Q(\mathbf{s},\mathbf{a}|\theta^Q)$ estimates the action value. The critic minimizes
\begin{equation}
\footnotesize
L(\theta^Q)=\frac{1}{B}\sum_{i=1}^{B}\big(y_i-Q(\mathbf{s}_i,\mathbf{a}_i|\theta^Q)\big)^2,
\label{eq:critic}
\end{equation}
with $y_i=r_i+\gamma Q'(\mathbf{s}_{i+1},\mu'(\mathbf{s}_{i+1}|\theta^{\mu'})|\theta^{Q'})$, where $\gamma$ is the discount factor. The actor follows the deterministic policy gradient:
\begin{equation}
\footnotesize
\nabla_{\theta^\mu}J\approx\frac{1}{B}\sum_{i=1}^{B}\nabla_{\mathbf{a}}Q(\mathbf{s}_i,\mathbf{a}|\theta^Q)\big|_{\mathbf{a}=\mu(\mathbf{s}_i)}\nabla_{\theta^\mu}\mu(\mathbf{s}_i|\theta^\mu).
\label{eq:actor}
\end{equation}

\subsection{Complexity Discussion}

AO repeatedly solves beamforming, scattering, and association subproblems, with complexity growing rapidly with $Q$, $F$, $M$, $N$, and $K_q$. In contrast, after offline training, DDPG requires only one actor forward pass and \rev{the closed-form action layer, whose cost is dominated by $F$ orthogonal completions of $M\times M$ matrices, i.e., $\mathcal{O}(FM^3)$ operations,} per slot, which suits online UAV operation.

\begin{figure*}[!t]
\centering
\begin{subfigure}{0.32\textwidth}
\centering
\includegraphics[width=\linewidth]{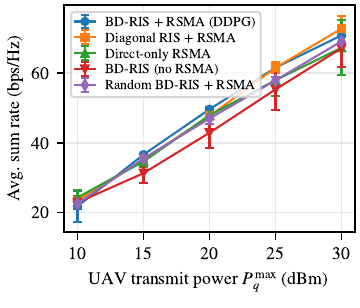}
\caption{Average network sum rate versus UAV transmit power.}
\label{fig:power}
\end{subfigure}
\hfill
\begin{subfigure}{0.32\textwidth}
\centering
\includegraphics[width=\linewidth]{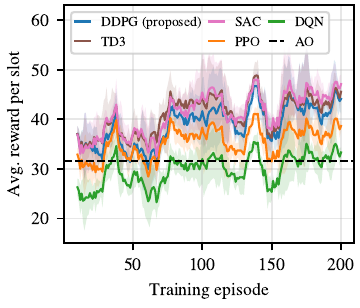}
\caption{Training convergence of the proposed and baseline algorithms.}
\label{fig:convergence}
\end{subfigure}
\hfill
\begin{subfigure}{0.32\textwidth}
\centering
\includegraphics[width=\linewidth]{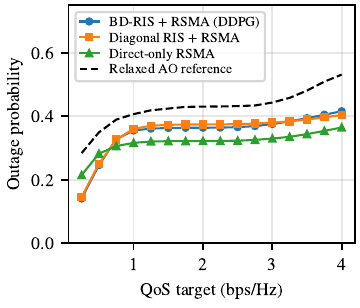}
\caption{Outage probability versus QoS target.}
\label{fig:outage}
\end{subfigure}
\caption{Performance of the proposed UAV-BD-RIS-RSMA framework.}
\label{fig:results}
\end{figure*}

\section{Numerical Results}

\begin{table}[!t]
\footnotesize
\centering
\caption{Simulation Parameters}
\label{tab:params}
\setlength{\tabcolsep}{2.2pt}
\begin{tabular}{@{}ll|ll@{}}
\toprule
\textbf{Parameter} & \textbf{Value} & \textbf{Parameter} & \textbf{Value} \\
\midrule
$(Q,F,K_q)$ & $(4,2,3)$ & $N$, $M$ & 8, 64 ($8\times8$) \\
$\sigma^2$, $W$ & $10^{-11}$ W, 1 MHz & $\beta_0$, $\alpha$, $\kappa$ & $-30$ dB, 2.2, 5 \\
Dense $a$, $b$, $\eta_{\mathrm{NLoS}}$ & 27.23, 0.08, 30 dB & Open $a$, $b$, $\eta_{\mathrm{NLoS}}$ & 4.88, 0.43, 20 dB \\
$V_q^{\max}$, $\Delta_t$ & 20 m/s, 1 s & Area & $1\times1$ km$^2$ \\
UAV altitude & 100--300 m & $R_{q,k}^{\min}$ & 1 bps/Hz \\
BD-RIS height & 25 m & $Z_0$ & 50 $\Omega$ \\
$\eta_1$, $\eta_2$ & 1, 0.1 & $F_q^{\max}$, $Q_f^{\max}$ & 1, 1 \\
$P_q^{\max}$ (train) & $\mathcal{U}[10,30]$ dBm & $P_q^{\max}$ (test) & 20 dBm \\
$E$, $T$ & 200, 100 & Replay, $B$ & $10^5$, 64 \\
Actor/critic & [400,300] & $\alpha_\mu$, $\alpha_Q$ & $10^{-4}$, $10^{-3}$ \\
$\gamma$, $\tau$ & 0.9, 0.005 & Noise $\sigma$, $\zeta$ & 0.1, 1 \\
Seeds, test ep. & 3, 10 & CSI error $\delta^2$ & 0.01, 0.05, 0.1 \\
\bottomrule
\end{tabular}
\end{table}

Table~\ref{tab:params} lists the simulation parameters. The BD-RISs are located at (250,250) and (750,750)~m. The UAVs fly from (200,200), (800,200), (200,800), and (800,800)~m to (800,800), (200,800), (800,200), and (200,200)~m, respectively, with an initial altitude of 150~m. The users are uniformly distributed in disks of radius 25~m centered at (285,285), (750,250), (250,750), and (715,715)~m, where clusters 1 and 4, next to the BD-RISs, are in the dense area. The reduced $(2,1,2)$ and large $(8,4,4)$ configurations are trained with $E=150$ and $E=120$, respectively. All algorithms are implemented in Python~3 with NumPy~2.4. Channels are regenerated at every slot according to the current UAV positions. Exploration noise is used only during training. To avoid tanh saturation, the actor loss of DDPG, TD3, and SAC includes the penalty $\zeta\|\mathbf{o}\|^2$ on the pre-tanh output $\mathbf{o}$. All results are averaged over 3 independent training seeds and 10 test episodes of $T$ slots and are reported as mean $\pm$ standard deviation.

\rev{\textit{Baselines:} i) \emph{BD-RIS without RSMA}: $\mathbf{w}_{q,c}=\mathbf{0}$, i.e., SDMA; ii) \emph{diagonal RIS + RSMA}: $\boldsymbol{\Phi}_f$ restricted to a diagonal unit-modulus matrix with phases $\angle y_{f,m}-\angle x_{f,m}$; iii) \emph{direct-only RSMA}: $b_{q,f}=0,\forall q,f$; iv) \emph{random BD-RIS}: random $\boldsymbol{\Psi}_f$ mapped by \eqref{eq:cayley}; v) \emph{DQN} \cite{mnih2015dqn}: five discrete levels per control variable (branching dueling architecture); vi) \emph{PPO} \cite{schulman2017ppo}: clipped on-policy policy gradient; vii) \emph{TD3}: twin critics with delayed policy updates; viii) \emph{SAC}: entropy-regularized off-policy learning; and ix) \emph{AO}: per-slot alternating optimization over association (greedy), BD-RIS (gradient ascent on $\boldsymbol{\Psi}_f$ through \eqref{eq:cayley}), and RSMA beamforming (projected gradient ascent), with straight-line UAV trajectories. All DRL methods use the same channels, state, action layer (DQN: discretized), network sizes, numbers of episodes, and test procedure; $P_q^{\max}$ is drawn from $\mathcal{U}[10,30]$~dBm per episode and included in the state, so one policy covers all power levels.}

\begin{table}[!t]
\footnotesize
\centering
\caption{Performance at $P_q^{\max}=20$ dBm (Mean $\pm$ Std over 3 Seeds)}
\label{tab:results}
\setlength{\tabcolsep}{12pt}
\begin{tabular}{@{}lccc@{}}
\toprule
\textbf{Scheme} & \textbf{Sum rate} & \textbf{Train} & \textbf{Inference} \\
 & \textbf{(bps/Hz)} & \textbf{(h)} & \textbf{(ms/slot)} \\
\midrule
\multicolumn{4}{@{}l}{\textit{Default $(4,2,3)$, perfect CSI}}\\
DDPG (proposed) & 49.6$\,\pm\,$0.6 & 0.08 & 1.4 \\
TD3 & 48.0$\,\pm\,$1.4 & 0.10 & 1.4 \\
SAC & 49.9$\,\pm\,$0.5 & 0.12 & 1.5 \\
PPO & 46.0$\,\pm\,$0.2 & 0.02 & 1.4 \\
DQN & 37.2$\,\pm\,$2.3 & 0.06 & 1.4 \\
AO & 34.1$\,\pm\,$0.4 & n/a & 91.1 \\
\midrule
\multicolumn{4}{@{}l}{\textit{Ablations for proposed design}}\\
w/o RSMA & 42.9$\,\pm\,$4.4 & 0.08 & 1.5 \\
Full $\boldsymbol{\Psi}_f$ (2080 outputs) & 41.2$\,\pm\,$0.9 & 0.28 & 1.4 \\
Nearest-BD-RIS assoc. & 48.9$\,\pm\,$3.1 & 0.08 & 1.5 \\
\midrule
\multicolumn{4}{@{}l}{\textit{Sensitivity and scalability}}\\
CSI error $\delta^2=0.01$ & 43.6$\,\pm\,$0.6 & -- & -- \\
CSI error $\delta^2=0.05$ & 34.6$\,\pm\,$0.7 & -- & -- \\
CSI error $\delta^2=0.1$ & 29.3$\,\pm\,$0.7 & -- & -- \\
Large $(8,4,4)$, DDPG & 87.3$\,\pm\,$2.1 & 0.09 & 2.9 \\
Large $(8,4,4)$, AO & 87.9$\,\pm\,$0.7 & n/a & 183.7 \\
\bottomrule
\end{tabular}
\end{table}

Fig.~\ref{fig:power} shows the average sum rate versus the UAV power budget. \rev{In Fig.~\ref{fig:results}, all curves show the mean over 3 seeds, with error bars or shading indicating $\pm$ one standard deviation; Figs.~\ref{fig:power} and \ref{fig:convergence} use the default configuration $(Q,F,K_q)=(4,2,3)$, and Fig.~\ref{fig:outage} uses the reduced configuration $(2,1,2)$, required by the relaxed AO reference. The sum rate of all schemes increases with the transmit power. RSMA provides the clearest gain: at 20~dBm, BD-RIS + RSMA achieves 49.6~bps/Hz, compared with 42.9~bps/Hz for BD-RIS without RSMA, because the common stream partially decodes the intra-cluster interference. BD-RIS + RSMA attains the highest sum rate at 15 and 20~dBm, about 2.0~bps/Hz higher than diagonal RIS + RSMA, 1.7~bps/Hz higher than direct-only RSMA, and 2.6~bps/Hz higher than random BD-RIS. These gains stem from the two blocked clusters and are modest relative to the seed-to-seed variability; at the other power levels, the RIS-assisted variants and direct-only RSMA are within about one standard deviation. This reflects the double path loss of the cascaded link, which is comparable to a blocked direct link but far weaker than an LoS one. Random BD-RIS does not outperform direct-only RSMA, i.e., the gain requires the CSI-aware scattering response, not merely beyond-diagonal coupling.}

\rev{In Fig.~\ref{fig:convergence}, where $P_q^{\max}\sim\mathcal{U}[10,30]$~dBm is drawn per episode and a 10-episode moving average is shown, the rewards rise during the first 100 episodes and then fluctuate around a plateau, since $P_q^{\max}$, users, and LoS states are redrawn in every episode. TD3 and SAC reach a slightly higher training reward than DDPG, PPO converges to a lower level, and DQN, which suffers from action discretization, is the lowest. The dashed line is the reward of per-slot AO, which is lower because AO keeps the straight-line trajectories.}

\rev{In Fig.~\ref{fig:outage}, obtained with $P_q^{\max}=20$~dBm, the outage saturates at a level set mainly by the blocked-cluster users. BD-RIS and diagonal RIS yield practically the same outage, and direct-only RSMA attains a lower outage above 0.5~bps/Hz, since its learned policy keeps the UAV over the blocked cluster for longer. The \emph{relaxed AO reference} is obtained by relaxing $b_{q,f}\in\{0,1\}$ to $b_{q,f}\in[0,1]$ and solving the resulting non-convex problem by AO with 5 random initializations. As AO only reaches a stationary point, this curve is a reference, not a guaranteed performance limit; with fixed trajectories, its outage is the highest.}

\rev{\textit{DRL baselines, ablations, and runtime:} Table~\ref{tab:results} shows that DDPG matches SAC (difference within one standard deviation) with about 35\% less training time and outperforms TD3, PPO, and DQN; hence, the gains stem mainly from the problem-specific design rather than from the RL algorithm. AO is 15.6~bps/Hz below DDPG, since it optimizes each slot for fixed straight-line trajectories, whereas DDPG also learns the trajectories. Removing RSMA reduces the sum rate by 6.8~bps/Hz. Replacing the closed-form BD-RIS layer by \eqref{eq:cayley} with $M(M+1)/2=2080$ actor outputs per BD-RIS reduces the sum rate by 8.5~bps/Hz (3 seeds, trained with $E=100$ owing to its computational cost) and increases the training time per episode 7.0-fold. Nearest-BD-RIS association is within one standard deviation of the learned one ($-0.7$~bps/Hz). Training time is measured over $E$ episodes, and inference time per action includes $\Gamma(\cdot)$. DDPG requires 5~min of training and 1.4~ms per slot, whereas AO needs 91~ms per slot.}

\rev{\textit{Imperfect CSI and scalability:} This sensitivity evaluation shows that, for $\delta^2=0.01$, $0.05$, and $0.1$, the sum rate decreases by 12\%, 30\%, and 41\%, respectively, owing to the interference leakage of the RZF beams of $\Gamma(\cdot)$ at high SNR. For the larger network $(8,4,4)$, trained with the same procedure and architecture, DDPG achieves 87.3$\,\pm\,$2.1~bps/Hz, comparable to AO (87.9~bps/Hz), with an about 64 times lower inference time. This case serves as an initial scalability stress test; a broader study, e.g., $(6,3,3)$ and $(10,5,5)$, is left for future work.}

\section{Conclusion}

We studied a BD-RIS-aided multi-UAV RSMA framework, in which only UAVs under favorable propagation conditions are assisted by BD-RISs, while the others use direct air-to-ground links. A joint sum-rate maximization over UAV trajectories, RSMA rate allocation, beamforming, reciprocal and lossless BD-RIS scattering matrices, and UAV-BD-RIS association was formulated and solved with the standard DDPG algorithm, tailored through a constraint-preserving action layer, a closed-form reciprocal and lossless BD-RIS response, and an environment-side association. 
Simulation results under a channel model with blocked direct links near the BD-RISs showed that RSMA provides the largest gain, that the CSI-aware BD-RIS response adds a modest gain at medium power, that DDPG matches SAC with less training time. Future work will address robust beam design under imperfect CSI and explore BD-RIS-assisted ISAC in integrated NTNs. 

\balance

\end{document}